\documentclass[ag]{copernicus}

\usepackage{mathptmx}

\usepackage{graphicx} 

\usepackage{amsmath}
\usepackage{amsfonts}
\usepackage{amssymb}
\usepackage{amsbsy}
\usepackage[figuresright]{rotating}
\usepackage{multicol}
\usepackage{array}
\usepackage{supertabular}

\usepackage{color}

\begin{document}

\title{{Magnetic field amplification in electron phase-space holes and related effects}}

\author[1,2]{{R. A. Treumann}
}
\author[3]{{W. Baumjohann}}

\affil[1]{Department of Geophysics and Environmental Sciences, Munich University, Munich, Germany}
\affil[2]{Department of Physics and Astronomy, Dartmouth College, Hanover NH 03755, USA}
\affil[3]{Space Research Institute, Austrian Academy of Sciences, Graz, Austria}

\runningtitle{Buneman two-stream instability}

\runningauthor{R. A. Treumann and W. Baumjohann}

\correspondence{R. A.Treumann\\ (rudolf.treumann@geophysik.uni-muenchen.de)}

\received{ }
\revised{ }
\accepted{ }
\published{ }


\firstpage{1}

\maketitle

\begin{abstract}
Three-dimensional electron phase space holes are shown to be positive charges on the plasma background which produce a radial electric field and force the trapped electron component into an azimuthal drift. In this way electron holes generate magnetic fields in the hole. We solve the cylindrical hole model exactly for the hole charge, electric potential and magnetic field. In electron holes, the magnetic field is amplified on the flux tube of the hole; equivalently,  in ion holes the field would be decreased. The flux tube adjacent to the electron hole is magnetically depleted by the external hole dipole field. This causes magnetic filamentation. It is also shown that holes are massive objects, each carrying a finite magnetic moment. Binary magnetic dipole interaction of these moments will cause alignment of the holes into chains along the magnetic field or, in the three-dimensional case, produce a magnetic fabric in the volume of hole formation. Since holes, in addition to being carriers of charges and magnetic moments, also have finite masses, they behave like quasi-particles, performing E$\times$B, magnetic field, and diamagnetic drifts. In an inhomogeneous magnetic field, their magnetic moments experience torque which causes nutation of the hole around the direction of the magnetic field presumably giving rise to low frequency magnetic  modulations like pulsations. A gas of many such holes may allow for a kinetic description in which holes undergo binary dipole interactions. This resembles the polymeric behaviour. Both magnetic field generation and magnetic structure formation is of interest in auroral, solar coronal and shock physics, in particular in the problem of magnetic field filamentation in relativistic foreshocks and cosmic ray acceleration. 

\end{abstract}

\introduction
\noindent
Within the last two decades, electron phase space holes have gained ever increasing importance in the dynamics of collisionless plasmas on the microscopic and also on the macroscopic scale. They cause small-scale structure in a hot collisionless plasma, particle acceleration and radiation, produce structure of the phase space distribution and also contribute to dissipation of free energy by heating plasmas and depleting energy which is stored in current flow, thereby providing conditions for other instabilities to evolve. Electron holes had been originally proposed by \citet{bernstein1957} as a nonlinear solution of the non-magnetised Vlasov-Maxwell system of equations which describes a hot collisionless electron plasma. Their statistical theory had been developed by \citet{dupree1972,dupree1982,dupree1983}. Fluid descriptions have identified electron holes as nonlinear solutions of the two-fluid dynamic equations of plasmas \citep{schamel1975,schamel1979,schamel1983, schamel1986,turikov1984}. Real progress on their structure and dynamics has, however, only been possible by properly designed numerical simulations \citep{newman2001,newman2002}. Still electron holes have remained an interesting theoretical concept of no further practical importance until spacecraft in situ observations in the near-Earth auroral environment provided a wealth of indications that electron holes do indeed exist in plasmas \citep{carlson1998,ergun1998a,ergun1998b,ergun1998c,pottelette2005}. Those observations have stimulated search for electron holes also in other space plasmas like Earth's bow shock, magnetopause, magnetotail and  in the solar wind. In several places their existence has indeed been confirmed. So, it is known now that electron holes are produced in the shock transition region from the solar wind to the magnetosheath \citep{bale1998a,bale2002}; evidence accumulated for their generation in the magnetopause during magnetic reconnection \citep{mozer2002,mozer2008,mozer2010,mozer2011}; the tail current sheet seems also to host electron holes in the presence of reconnection and under other conditions \citep{pickett2005}. They have been found in the magnetosheath \citep{vaivads2004} possibly also in relation to reconnection in the polar cusp. Sophisticated numerical simulations have meanwhile investigated their configuration and phase space properties in one and two dimensions. Simulations of reconnection \citep{drake2004} under prescribed conditions have also been claimed to show the evolution of electron holes along the separatrices.

All these observations and theories suggest that electron holes or, in a wider sense, phase space holes including electron and ion holes, should play an important role in astrophysical plasmas as well. The problem about those remote systems is that no in situ observations are possible to perform. One single such measurement would probably completely change our attitude to the behaviour of gaseous matter under  astrophysical conditions shifting all our current fluid dynamical theories into the domain of kinetic theory. This has partially been done already in relativistic shock theory where some observations force one to include kinetic aspects \citep[for a recent critical review see, e.g.,][]{bykov2011}. However, the role of structure on the scale of phase space holes has so far not been considered to be of any importance. The example of space plasma probably proves such an attitude premature. We have recently shown that electron holes seem to be strong sources of non-thermal radio radiation from the aurora \citep{treumann2012} and with high probability also in other planetary objects. 

\section{Extension of the Buneman instability}
In view of application to the various regions where electron holes are expected to be generated, i.e. the auroral regions of magnetised planets, active regions in the solar corona, the source regions of solar flares, collisionless shock transitions, and relativistic shocks as sources of cosmic ray acceleration, it makes sense to briefly summarise the mechanism of electron hole production, before developing the analytical theory of magnetic field generation in electron holes. This is done in the present section in a way which generalises the main ingredients of this mechanism to relativistic plasmas.

 Phase space holes evolve from plasma instability. The most important instability with respect to electron holes is the Buneman two-stream instability \citep{buneman1958,buneman1959}. This is a low-frequency electrostatic instability generating a strong longitudinal wave electric field. It is sometimes claimed that electron holes would also be generated by the electron-acoustic instability. However, careful numerical simulations \citep{matsukiyo2004} have unambiguously shown that this is barely the case. Electron-acoustic waves evolve in two temperature and/or counter streaming electron plasmas. Since they are highly Landau damped, other instabilities like the electron two-stream (Langmuir) instability grow faster and dominate the spectrum from the beginning, even under extreme conditions, with electron acoustic waves being produced only at a very weak wave level incapable of evolution to large enough amplitudes for hole formation.

The Buneman mode evolves along the magnetic field and propagates with phase velocity $v_B<V_d$ less but close to the bulk velocity difference, $V_d= V_i-V_e$, between electrons and ions. It is a \emph{current driven} instability. It is therefore important in all cases when the plasma carries a strong current\footnote{This fact raises strong doubts on some very strong claims that the Buneman instability arises as a competitor to the Weibel mode in generally current neutral electron beam-beam interactions \citep{bret2009,bret2011}. The best investigation of the complete transition from the two-stream to the Buneman mode has been published long ago \citep{dum1989,dum1994}. It goes far beyond the above papers. In order to drive the Buneman mode, a mechanism is needed which maintains the \emph{parallel} current flow. Such mechanisms are provided in the auroral region and, in relation to relativistic shocks, only in the Bell cosmic ray shock regime where sufficiently strong field-aligned cosmic ray return currents flow in the relativistic foreshock background plasma. For \emph{perpendicular} current flow like in (relativistic and non-relativistic) shock transitions, however, the Buneman mode is stable. It is replaced there by the modified two-stream instability which has completely different properties. In the above papers the waves misleadingly claimed to be Buneman are at the best weak electron acoustic modes which are incapable of hole formation and play no role in plasma dynamics.}. Its fluid-like nature implies that the current must be strong in the sense that the electrons can be considered cold, i.e. their thermal speed, $v_e<V_d$, must be less than the current drift velocity. The smallness of $v_e$ permits the use of the fluid approximation, neglecting thermal effects in order to find the most pronounced reactive instability of the plasma. However, only the inclusion of thermal effects is responsible for hole formation during nonlinear evolution of the instability, as noted above. On the other hand, large thermal speeds and related effects suppress the Buneman instability  replacing it by the purely kinetic ion-acoustic instability which favours ion holes to evolve nonlinearly instead of electron holes.

Viewed from the ion frame, one has $V_d=-V_e$ with $V_e<0$.  Then, in the simplified non-relativistic case, the Buneman dispersion relation becomes \citep{baumjohann2012}
\begin{equation}
D(\omega, k)=1-\frac{\omega_i^2}{\omega^2}-\frac{\omega_e^2}{(\omega-kV_d)^2}=0
\end{equation}
It is a fourth order equation for the frequency $\omega$. Of the solutions, the high frequency mode $\omega\approx\omega_e+kV_d$, which propagates on the Langmuir wave branch, is of no interest and is not excited. It decouples from the three other modes, which have wave number $k\approx \omega_e/V_d$ and thus much lower frequency, leaving for them the dispersion relation
\begin{equation}
\omega^3\approx -\frac{m_e}{2m_i}\omega_e^3
\end{equation}
The frequency of these waves is reduced by the electron to ion mass ratio on the right. The complex solutions to this dispersion relation are easily found putting $\omega=\omega_B+i\gamma_B$, with $\omega_B$ the real frequency, and $\gamma_B$ growth rate. The unstable Buneman wave solution has frequency and growth rate
\begin{equation}
\omega_B\approx\omega_e\left(\frac{m_e}{16m_i}\right)^\frac{1}{3}, \qquad \gamma_B=(3)^\frac{1}{3}\omega_B
\end{equation}
being a slightly faster growing than oscillating reactive wave in which the entire electron distribution participates. Just for this reason it preferentially causes electron holes and does not simply stabilise by depleting the current. This is easy to understand. The Buneman instability is not simply an absolute instability; it also grows convectively. This cannot be seen from the above formulae. It requires a somewhat more sophisticated approach to deduce the convective growth rate
\begin{equation}
\mathrm{Im}\ k=-\frac{\omega_B}{\partial\omega/\partial k}\approx -\frac{\gamma_B}{V_d}\left(\frac{16m_i}{m_e}\right)^\frac{1}{3}
\end{equation}
The waves thus grow along their direction of propagation with the electrons.

Since the whole electron population is affected by the instability and starts oscillating, a violent bulk plasma motion back and forth is induced by the growing Buneman mode longitudinal electric field. At the same time the mode moves with the electrons at a smaller though substantial (phase and group) velocity 
\begin{equation}
v_B\equiv \frac{\omega_B}{k}\approx V_d\left(\frac{m_e}{16m_i}\right)^\frac{1}{3}\approx \frac{\partial\omega}{\partial k}
\end{equation}
The electrons are, of course, not cold but have a substantial velocity spread, $v_e<V_d$. In the wave frame, depending on the phase of the electric field oscillation, the electric potential field $\Phi$ traps the low electron energy $\epsilon_e=\frac{1}{2}m_ev^2<e\Phi$ component of the electron distribution in the negative potential valleys when passing the wave; from the positive potential wells these electrons are expelled. Higher energy electrons pass the potential troughs and wells and become accelerated. 

\begin{figure}[t!]
\centerline{{\includegraphics[width=0.5\textwidth,clip=]{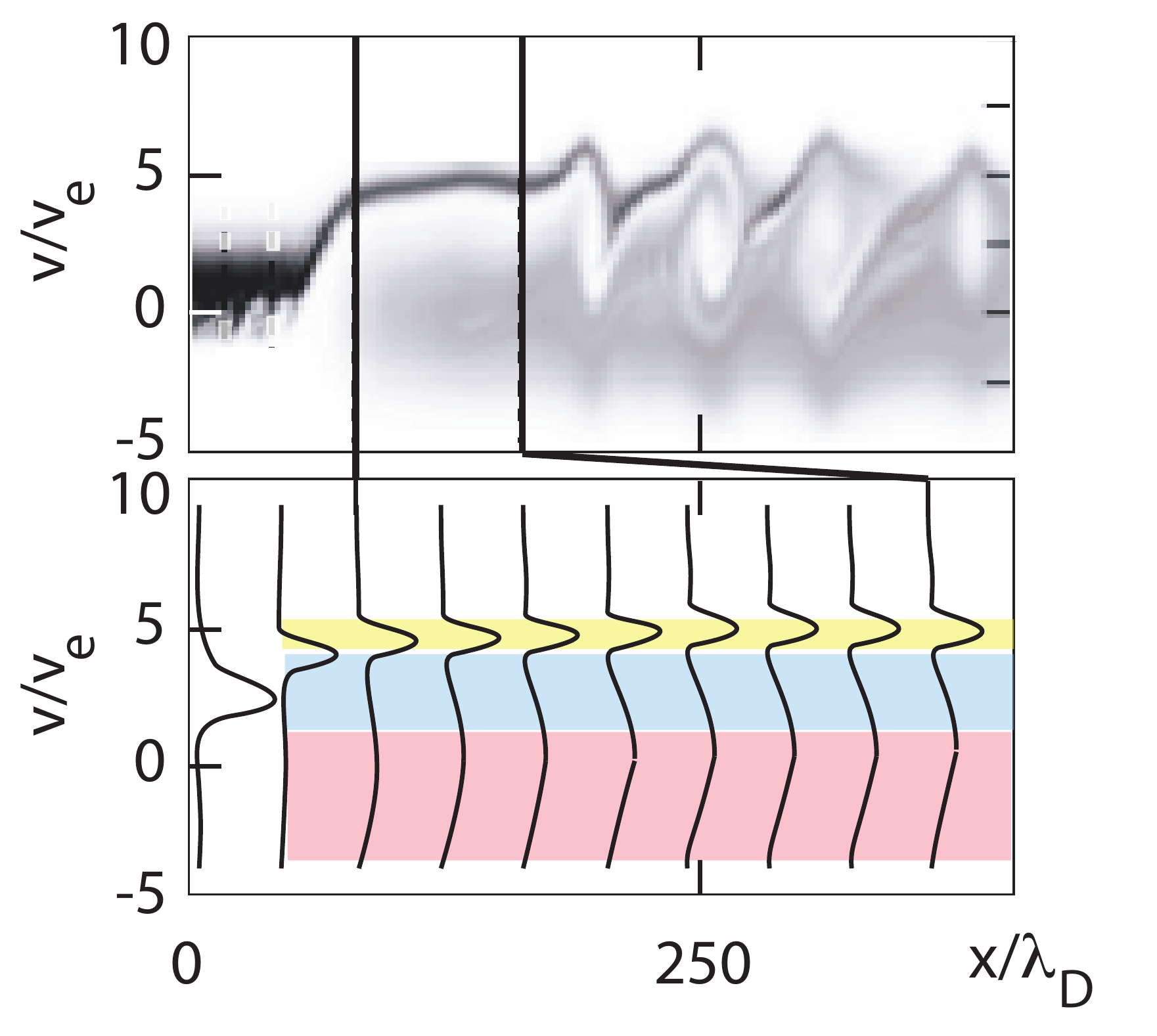}
}}\vspace{-3mm}
\caption[ ]
{\footnotesize {An example of electron hole formation in a one-dimensional simulations \citep[data from][]{newman2002}. The upper panel shows the phase space, $(x,v)$, of electrons which are injected as a warm (thermal speed $v_e$) current carrying electron beam on the left, satisfying the Buneman condition. Velocity is scaled in terms of the thermal speed, $v_e$, space in terms of the debye length, $\lambda_D=v_e/\omega_e$. Evolution into a cold electron beam at velocity $v\approx 5 v_e$ is readily seen in relation to the generation of phase space holes from where electrons are emptied. Parallel to beam evolution a hot electron component is generated by trapping and heating. Hole structure is not simply dipolar as suggested by theory. Instead, negative velocities indicate reflection of electrons and complicated holes, consisting of several highly dynamical electron layers of low phase space density. The lower panel shows average distribution cuts at different locations in space. Indicated yellow is the cold accelerated beam, blue the hole region which is about void of electrons, red the warm hole-trapped electron population. Spatial hole extension is several ten Debye lengths. The phase-space extension of the hole is of the order of $\lesssim 10 v_e$, substantially wider than the thermal width ($\sim 2v_e$) of the initial beam.}}
\label{fig-bun-1}
\end{figure}

During wave growth, this process causes ever deeper localised potential wells containing ever less electrons. This leads to the formation of a whole chain of holes and, at the same time, a cold accelerated electron beam consisting of the fastest electrons, as shown in the schematic of simulation data in Fig. \ref{fig-bun-1}. This process is highly nonlinear and depends on the electron and wave potential phase and the mutual  interaction among the electrons. 

The cold electron beam still carries current and remains unaffected by the holes except for some quasi-periodic energy modulation which is due to its passing the sequence of holes. Ions do not participate in this dynamics because the frequency $\omega_B$ and growth rate are substantially far above the ion plasma frequency $\omega_i$. Of course, in the late stage of hole evolution when the electric field in the holes is quasi-stationary, ions become affected as well, mainly by the evolution of  density gradients on the boundaries of the many electron holes thus transforming the homogeneous plasma into an inhomogeneous conglomeration of density fluctuations.   

The obvious effects of the presence of Buneman generated electron holes are therefore 

-- the separation of the electron distribution into a warm trapped component inside the holes and a fast cold electron beam carrying the residual current; 

-- the production of a spatial density structure which appears as a quasi-periodic modulation along the magnetic field;  

-- the generation of localised electrostatic potentials which are organised  into a chain of negative potential wells along the current flow;

-- the production of the cold accelerated electron beam is of particular interest. It is the result of a very interesting mechanism of non-radiation electron cooling provided by electron holes, the prospects and consequences of which for application have not yet been explored, neither in the laboratory nor in view of astrophysical application. This electron cooling which maintains the accelerated cold beam is much more efficient than any electron cooling by radiation can provide when considering that radiation cooling is a higher order effect. 

Implicitly these results were contained already in the work of \citet{bernstein1957}, but they have been elucidated only by sophisticated and carefully performed numerical simulations.

In application to astrophysics one needs to switch to a relativistic formulation of the Buneman mode and electron hole formation. Since astrophysical current speeds (for instance in a relativistic foreshock populated with cosmic rays of high Lorentz factor) are close to the velocity of light, one expects very strong effects. Assuming current speeds $V_d\approx c$ but nonrelativistic temperatures and thermal speeds, $v_e, v_i\ll c$, we may transform the above theory into a relativistic theory. In this case the plasma frequencies are unaffected by the bulk motion, leaving $\omega_i, \omega_e$ the non-relativistic plasma frequencies. A bulk Lorentz factor $\Gamma$ of the electrons cancels out of the relativistic expressions. On the other hand, for finitely large relativistic temperatures the electron plasma frequency would have to be defined as
\begin{equation}
\omega_{e,rel}=\omega_{e}\int d^3\!\!\!\!p\ F_e(p)/\gamma_e(p)
\end{equation}
where $p$ is the proper individual relativistic particle momentum, $F_e(p)$ the relativistic electron distribution (for instance the J\"uttner distribution), and $1\leq\gamma_e(p)<\infty$ is the internal electron Lorentz factor. This expression shows that it is essentially the low energy electrons which determine the plasma frequency, in particular when the distribution drops steeply with increasing $p$. Hence, one may safely use the non-relativistic plasma frequency only. 

The above dispersion relation for the Buneman instability holds in the non-relativistic ion frame. In this frame, in the relativistic case, the velocity becomes $V_d=c\beta$, and one must replace the frequency by its Doppler shifted analogue. Moreover, the wave number $k\to k\Gamma$ varies due to Lorentz contraction. Then for sufficiently large bulk $\Gamma$ factors, we have $V_d\approx c$ and can make the following replacements:
\begin{equation}
\omega\to\omega/\sqrt{2\Gamma}, \qquad kV_d\to kc\Gamma 
\end{equation}
This yields the Buneman wavenumber, $k\approx \omega_e/c\Gamma$, and dispersion relation,
\begin{equation}
\frac{\omega^3}{\omega_e^3}\approx-\frac{m_e}{2m_i}\left(2\Gamma\right)^{-\frac{9}{2}}
\end{equation}
which has the growing solution
\begin{equation}
\omega_{B,rel}\approx\left(\frac{m_e}{16m_i}\right)^\frac{1}{3}\frac{\omega_e}{\sqrt{8\Gamma^3}}, \quad \gamma_{B,rel}\approx (3)^\frac{1}{3}\omega_{B,rel}
\end{equation}
As expected, time dilatation slows the Buneman instability down with increasing relativistic factor $\Gamma$. Nevertheless, the Buneman instability exists in this case, grows and, in the same way as in non-relativistic plasmas, causes the generation of electron holes, depletes the electron density from positive electric potential wells and otherwise traps electrons of energy 
\begin{equation}
\epsilon_e=m_ec^2\gamma_e(p)\leq e\Phi 
\end{equation}
in negative potential wells, $\Phi$. This is probable because, with increasing bulk Lorentz factor, the effect of the thermal velocity on the instability decreases. 

The Buneman instability thus gains importance in ultra-relativistic current carrying flows. In contrast to the Weibel instability \citep{weibel1959}, the behaviour of the instability is primarily not affected by the presence of a magnetic field, whether weak or moderately strong, as long as the current flows parallel to the field.  Such field-aligned `return currents' are central to Bell-type models of magnetic field amplification in the vicinity of relativistic shocks and are assumed to work in the relativistic shock foreshock for nearly parallel shocks \citep{bell2004,bell2005,bykov2011a} in the presence of Cosmic Rays and shock acceleration of Cosmic Rays. The idea in these models is that the Cosmic Ray return currents cause a non-resonant instability to grow which is believed to substantially increase the foreshock magnetic field. However, these currents, if present, should in the first place excite the Buneman mode long before pushing the Bell mode to grow.

In the following we show that electron holes excited by the Buneman mode along an ambient magnetic field are themselves efficient magnetic field generators on a much faster time scale than the Bell modes. The field-generation time scale is comparable to the Buneman formation time of electron holes, which is of the order of $\Delta t_h\sim 10^3 \omega_{e}^{-1}$; compared to the Bell mode this is instantaneous.  However, the spatial scales of the holes are small as well. Along the magnetic field they scale with the Debye length as  $\sim 10^2\lambda_D$, transverse to the magnetic field holes may be oblate or narrow.  Here their scales are of the order of the hole-trapped electron population gyroradius, which in a relativistic plasma is of the order of the inertial scale $\sim c/\omega_e=\lambda_e$, comparable to the scale of the Weibel instability. Thus, being generators of magnetic fields, electron holes structure the magnetic field on these scales. 

The formation of holes in all details cannot be described analytically. It requires the use of numerical simulations, as has been done for the non-relativistic Buneman instability \citep[see, e.g.,][and other less sophisticated simulation studies]{muschietti1999a,muschietti1999b,newman2001,newman2002}.

\section{Generation of magnetic fields}
Electron holes may generate electromagnetic waves propagating on the whistler and Z-mode branches \citep{bale1998b}. The former have been known as saucer emissions in strong magnetic fields \citep{newman2002, oppenheim2001,ergun2002a,ergun2002b,du2011,wu2011}, the latter may result from some electron cyclotron maser instability in much weaker magnetic fields \citep{treumann2006,treumann2012}. 

The presence of magnetic fluctuations and quasi-stationary magnetic fields in relation to the generation of electron phase space holes has recently been confirmed by {\small THEMIS} observations \citep{andersson2009,tao2011}. 

The original idea for magnetic field generation in holes started from an electron hole being a positive charge on the electron background and aimed at the interaction of such charges for production of bremsstrahlung radiation. Subsequently, simulations have been performed showing that holes indeed possess magnetic modes \citep{terry1990,jovanovic1993,oppenheim2001,newman2002,chen2002,chen2004,umeda2004,umeda2006,du2011,wu2011} with the main interest being on the generation of low frequency electromagnetic waves focussing on the stability of holes with respect to the radiation of whistlers, the production of saucer emissions, and the decay of phase space holes. Here, we treat an analytical model of an electron hole with emphasis on the formation of quasi-stationary magnetic fields and their consequences.

\subsection{\small{Preliminaries: Model settings}}
\begin{figure}[t!]
\centerline{{\includegraphics[width=0.5\textwidth,clip=]{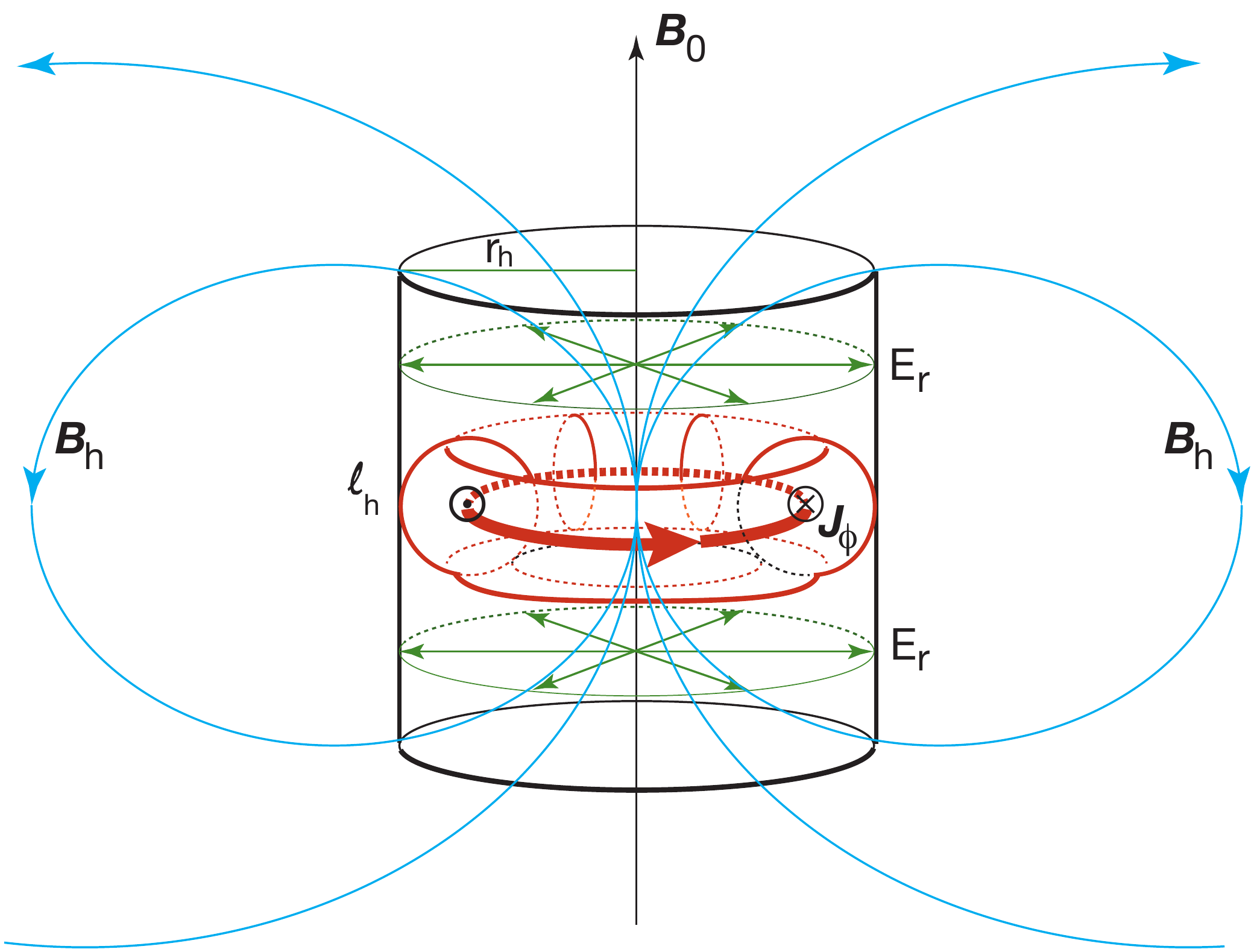}
}}\vspace{-3mm}
\caption[ ]
{\footnotesize {The cylindrical model of the electron hole used in the calculation. The hole has length $\ell_h$ along the ambient magnetic field $\mathbf{B}_0$ and radius $r_h$. Its positive charge generates the radial electric field $E_r$ shown here in green colour. Electron perform an E$\times$B drift in the crossed fields $E_r$ and $B_0$ which is in $-\phi$ direction. This drift corresponds to a current  $\mathbf{J}$ direction in $+\phi$. This solenoidal current, shown in red, generates the hole magnetic field which outside the hole appears as the dipolar magnetic field $\mathbf{B}_h$ of which here only one planar section is shown in blue. }}
\label{fig-bun-4}
\end{figure}

Assuming that, in a very simplified model, the electron hole has cylindrical shape of radius $r_h$ and length $\ell_h$ along the magnetic field,\footnote{There is no consensus about the shape of electron phase space holes. \citep[For theoretically and observationally inferred shapes of holes see, e.g.,][]{muschietti1999a,franz2000}. Theory and simulations are not developed far enough for unambiguously determining the canonical shape of holes in real and in phase space. Observations, on the other hand, suggest either circular, elongated or oblate shapes. For simplicity we restrict to a cylindrical form of the holes, here.} then it contains, in addition to its parallel electric hole field, $E_z$, a radial electric field, $\mathbf{E}_\perp=E_\perp \hat r$, which results from the depleted (missing) electrons in the hole core as compared to the hole environment. The model is illustrated in Fig. \ref{fig-bun-4}. This corresponds to a \emph{positive} line charge distributed over the full length of the electron holes. In general, this hole-forced space charge density $\rho_h=-e\Delta N>0$ is positive because the electron density is depleted inside the hole:  
\begin{equation}
\Delta N=N_h(r,z)-N<0
\end{equation}
$N_h$ is a complicated function of the spatial coordinates, $z,r$, which is determined by the nonlinear dynamics of hole formation and also varies with time, $t$. In simplified models one assumes it to be constant. In the electron hole the radial electric field related to $\rho_h$ is directed radially outward. [In ion holes, effectively representing negative space charges, the field would be directed inward, instead.] The total positive space charge carried by the electron hole is obtained from
\begin{equation}\label{eq-holecharge}
q_h=-2\pi e\int\limits_0^{r_h} rdr\int\limits_{-\ell_h/2}^{\ell_h/2}dz\ \Delta N(r,z)
\end{equation}

Assuming that the trapped hole electron gyroradius $r_{ec}<r_h$ is sufficiently small against the hole radius, $r_h$, then the trapped electrons are magnetised and will perform an electric field drift in the combined hole-electric $E_r$, and ambient, $\mathbf{B}_0=B_0\,\hat z$, magnetic fields. Untrapped electrons of the fast accelerated electron beam mentioned previously do not contribute to this drift. The above conditions also imply a limit on the trapped electron energy 
\begin{equation}
\epsilon_t={\textstyle\frac{1}{2}}m_ev^2 \lesssim {\textstyle\frac{1}{2}}\frac{m_er_h^2}{\omega_{ce}}
\end{equation}
The corresponding relativistic expression is
\begin{equation}
\epsilon_t\lesssim m_ec^2\sqrt{r_h/\lambda_e} 
\end{equation}
where the proper relativistic Lorentz factor has been replaced through its relation to the relativistic gyroradius, $r_{ce}=\lambda_e\gamma_e^2$. This gives rise to a clockwise azimuthal electron drift in the crossed positive radial electric and magnetic fields, 
\begin{equation}
 V_ \phi =E_r/B_0 
\end{equation}
which creates an anti-clockwise azimuthal electric current 
\begin{equation}
J_\phi=-\frac{eE_r(r,z)}{B_0}N_h(r,z)>0
\end{equation}
Only the hole-trapped electrons participate in this drift and contribute to the current. The electric field is derived from a scalar potential, $\Phi_h$, as $\mathbf{E}=-\nabla\Phi_h$ with radial component $E_r=-\partial\Phi_h/\partial r$.

The magnetic field, $\mathbf{B}$, generated by the solenoidal current, $J_\phi$, flowing inside the hole is a dipolar field, when seen from the outside. Its strength can easily be estimated if the hole current is known. Since neither the current density nor density distribution are know with sufficient precision, any sophisticated model calculation does not provide further insight. It is completely sufficient to use the well-known expressions for the magnetic field of a cylindrical current loop (or solenoidal current). Outside the hole  the magnetic field generated by the total azimuthal  hole-electric current, $I_\phi$, becomes
\begin{equation}
\mathbf{B}_\mathit{ex}= \frac{\mu_0I_\phi}{2R^3}\big(\cos\theta, 0, {\textstyle\frac{1}{2}}\sin\theta\big), \qquad R=\frac{r}{r_h} \gg 1 
\end{equation}
which is the expected dipole field. Inside the hole at $R<1$ the field has components $B_r, B_\theta$, given by
\begin{equation}\label{eq-bfield}
\Bigg(
\begin{array}{c}
  B_{hr}     \\ [0.8ex]
   B_{h\theta}       
\end{array}
\Bigg)=\frac{\mu_0 I_\phi}{4r_h}\frac{\big(2+R^2+R\sin\theta\big)}{\big(1+R^2+2R\sin\theta\big)^{5/2}}
\Bigg(
\begin{array}{r}
  \cos\,\theta  \\ [0.8ex]
 - \sin\,\theta     
\end{array}
\Bigg)
\end{equation}
which, for $\theta=0$, on the axis degenerates into the component $B_z$ along the $z$-axis. In this simplified model it suffices to use the value 
\begin{equation}
B_z(0.0)=\mu_0 I_\phi/2r_h\quad\mathrm{at}\quad R=0,\theta=0
\end{equation}
for the internal field and to determine the total hole-electric current by integrating over the surface crossed by the azimuthal current, $J_\phi (r,z)$, according to
\begin{equation}\label{eq-currenti}
I_\phi =\Bigg|\frac{e}{B_0}\int\limits_0^{r_h} dr\int\limits_{-\ell_h/2}^{+\ell_h/2} dz\, E_r(r,z)N_h(r,z)\Bigg|
\end{equation}

This secondary hole-generated magnetic field is along the original field and thus \emph{amplifies} the ambient magnetic field inside the hole over its length $\ell_h$. Being dipolar outside the hole, the hole magnetic field weakens  the external field and, at the same time, exerts a repulsive force on the next hole neighbour. 

\subsection{\small{Hole potential boundary value problem}}
In order to infer about the amplification one needs to calculate the current strength $I_\phi$. In principle this includes the complete dynamics of electron hole generation, determination of the hole density $N_h$ and solution of the Poisson equation for the electric field. Numerical calculations of this kind using cylindrical hole models and assuming Gaussian electric potential fields have been performed by \citet{chen2002}. They do not lead to deeper insight and are thus of limited value. Rough estimates can be obtained assuming that the hole is either elongated along the magnetic field with $\ell_h\gg r_h$ or is oblate with $\ell_h\ll r_h$. Observations \citep{andersson2009} suggest that the former case is more probable. Then, the Poisson equation for the electric potential $\Phi_h$ simplifies to
\begin{equation}
\nabla^2 \Phi_h (r,z)\approx -\frac{e}{\epsilon_0}\big(N-N_h(r)\big)
\end{equation}
The hole density can be modelled, assuming that the density is depleted along the magnetic field by the parallel electric field. This assumption is appropriate for the case of magnetised electrons. These are confined to the magnetic field and can escape only along the magnetic field. Restriction to the action of the parallel field becomes problematic in the late stages of magnetic field generation when the amplified internal magnetic field contributes to squeezing electrons out of the hole via the mirror force. Neglecting this effect we have
\begin{equation}
N_h=N\exp \left[\frac{e\Phi_h}{T_{eh}}\right]\approx N\left(1+\frac{e\Phi_h}{T_{eh}}\right)
\end{equation}
where $T_e$ is the electron temperature of the electrons. Inserting into Poisson's equation yields
\begin{equation}
\nabla^2\Phi_h(r,z)\approx \frac{e^2N}{\epsilon_0T_{eh}}\Phi_h(r,z)=\frac{\Phi_h(r,z)}{\lambda_{Dh}^2}
\end{equation}
where we introduced the hole-Debye length, $\lambda_{Dh}$, allowing for a different hole temperature. This equation which must be solved for the interior of the hole and glued to the external solutions via the boundary conditions at the hole surface. Since this is impossible to do for a real hole, we choose the above described idealised cylindrical hole model of perpendicular hole radius, $r_h$, and parallel extension $\ell_h$. 

In the outer hole space the space charge does not vanish completely, however. The field can penetrate the external region a certain distance, either the electron skin depth $\lambda_e$ or the external Debye length $\lambda_D$. Since the hole is a positive charge, it is expected that it attracts some electrons to accumulate them near the outer hole boundary over the distance of field penetration. Hence, $\Delta N\neq 0$ outside close to the hole, and the Poisson equation for the external potential $\Phi_{ex}^0$ does not degenerate to the Laplace equation but reads
\begin{equation}
\nabla^2\Phi_\mathit{ex}^0(r,z)=-\frac{\rho_{ex}}{\epsilon_0},\quad\mathrm{with}\quad \rho_{ex}=-eN\exp\left[\frac{e\Phi_\mathit{ex}^0}{T_e}\right]
\end{equation}
for the external potential $\Phi_\mathit{ex}(r,z)=\Phi_\mathit{ex}^0+T_e/e$. This yields the external Poisson equation
\begin{equation}
\nabla^2\Phi_\mathit{ex}(r,z)=\Phi_\mathit{ex}(r,z)/\lambda_D^2
\end{equation}

The boundary conditions to be satisfied concern the continuity of the tangential electric field components at the boundaries of the cylindrical hole. These imply the continuity of the field aligned electric field $E_z\big|_{r_h}=-\partial\Phi/\partial z\big|_{r_h}$ at the cylindrical radial surface $r=r_h$, and continuity of the radial electric field $E_r\big|_{\ell_h/2}=-\partial\Phi/\partial r \big|_{\ell_h/2}$ at the top and bottom surfaces of the hole at $z=\pm\ell_h/2$
\begin{eqnarray}
~~~~~~~~~~E_{hz}(z)&=&E_\mathit{ex,z}(z) \quad\mathrm{at}\quad r=r_h \\
~~~~~~~~E_{hr}(r)&=&E_\mathit{ex,r}(r) \quad\mathrm{at}\quad z=\pm\ell_h/2
\end{eqnarray}
Poisson's equation for the hole potential in cylindrical coordinates reads
\begin{equation}
\frac{1}{r}\frac{\partial}{\partial r}r\frac{\partial\Phi_h(r,z)}{\partial r} +\frac{\partial^2\Phi_h(r,z)}{\partial z^2}\approx \frac{\Phi_h(r,z)}{\lambda_{Dh}^2}
\end{equation}
A similar equation holds for the external potential $\Phi_\mathit{ex}$. These equations can be solved by separation of variables. The second boundary condition can be taken care of by letting the radial electric field vanish at the top and bottom of the hole. In addition one needs to make sure that the fields vanish at infinity, $r\to\infty,z\to\pm\infty$. With all these conditions the final solutions for the potentials become
\begin{eqnarray}
\Phi_h(r,z)&=&\Phi_0I_0\left(x\right)\cos\,\left(\frac{\pi z}{\ell_h} \right), \qquad\qquad  ~~~|z|\leq\frac{\ell}{2} \label{eq-phihole} \\
\Phi^0_\mathit{ex}(r,z)&=&\Phi_0\frac{I_0\left(x_h\right)K_0\left(x'\right)}{K_0\left(x'_h\right)}\cos\left(\frac{\pi z}{\ell_h}\right),\quad  ~~|z|\leq\frac{\ell}{2} \\
\Phi^0_\mathit{ex}(r,z)&=&\Phi_0\frac{I_0\left(x_h\right)K_0\left(x'\right)}{K_0\left(x'_h\right)} g(z), \qquad\qquad\, |z|\geq\frac{\ell}{2} 
\end{eqnarray}
where the  functions $I_0(x),K_0(x)$ are modified Bessel functions, and we defined
\begin{equation}
x=: \frac{r}{\lambda_{Dh}}\left[1+\left(\frac{\pi\lambda_{Dh}}{\ell_h}\right)^2\right]^\frac{1}{2}\!\!\!\!, \  x'=: \frac{r}{\lambda_D}\left[1+\left(\frac{\pi\lambda_{D}}{\ell_h}\right)^2\right]^\frac{1}{2}
\end{equation}
and $x_h,x'_h$ are taken at $r=r_h$. The function $g(z)$ takes care of the decay of the electric field in $\pm z$ direction. Its specific form depends on the plasma outside the hole. Since the potentials vanish at $z=\pm \ell_h$ throughout all space, it depends on the penetration depth of the fields. Though it will not be important for the following discussion, a simple expression can be given as 
\begin{equation}
g(z)=\left\{ 
\begin{array}{lccccr}
 1-  \exp \left[-\alpha\left(z-\frac{1}{2} \ell_h\right)\right],& &&& & z- \frac{1}{2}\ell_h\geq0   \\ [1.3ex]
 1-  \exp \left[-\alpha\left(\frac{1}{2}\ell_h-z\right)\right],&  &&& &z+\frac{1}{2}\ell_h\leq0   
\end{array}
\right.
\end{equation}
where either $\alpha=\lambda_D^{-1}$ or $\alpha=\lambda_e^{-1}$  is the inverse of the damping length of the electric field outside the hole.

These expressions hold for both elongated and oblate electron holes. In all cases the physics of the holes tells that the elongation along the magnetic field exceeds the Debye length such that the expression under the root will simplify. Moreover, $\Phi_0$ is the extremum of the electric potential on the axis of the hole. Since the electric field points outward, we have $\Phi_0=-U<0$.

\subsection{\small{Radial hole electric field and azimuthal current $I_\phi$}}
From the above expressions for the potentials it is straightforward to calculate the radial electric field $E_r$ in the hole for insertion into the drift velocity $V_\phi$:
\begin{eqnarray}
~~~~~~~~~~E_r(r,z)&=&E_{r0}I_1\left(x\right)\cos\left(\frac{\pi z}{\ell_h}\right) \\
E_{r0}&\equiv& \frac{U}{\lambda_{Dh}}\left[1+\left(\frac{\pi\lambda_{Dh}}{\ell_h}\right)^2\right]^\frac{1}{2}
\end{eqnarray}
This yields for the current density in the interior of the electron hole
\begin{equation}
\mathbf{J}= J_\phi(x,z)\hat\phi\approx \frac{\epsilon_0U\,E_{r0}}{\lambda_{Dh}^2B_0}I_0(x)I_1(x)\cos^2\left(\frac{\pi z}{\ell_h}\right) \hat\phi
\end{equation}
The current strength in the equivalent loop representing the hole-generated current is obtained by integration from Eq. (\ref{eq-currenti}). This integration can be performed quite generally, observing that $E_r=-\partial\Phi_h/\partial r$, and $N_h=N\exp(e\Phi_h/T_{eh})$. The $r$-integrand becomes a total derivative, leaving us with 
\begin{equation}
I_\phi=\frac{NT_{eh}\ell_h}{\pi B_0}\int\limits_{-\pi/2}^{\pi/2} d\zeta\left\{\mathrm{e}^{-s\cos\,\zeta}-\mathrm{e}^{-sI_0(x_h)\cos\,\zeta}\right\}
\end{equation}
where $s=eU/T_{eh}$, and $\zeta=\pi z/\ell_h$ is a dummy integration variable. This yields for the current
\begin{equation}
I_\phi=\frac{NT_{eh}\ell_h}{B_0}\Big\{I_0\big[sI_0(x_h)\big] -I_0(s)\Big\}
\end{equation}
Expanding the Bessel functions to second order yields an approximate but more useful expression for the current in the hole:
\begin{equation}
I_\phi\approx\frac{B_0\ell_h}{4\pi\mu_0} \frac{\big[I_0(x_h)\big]^2-1}{1+(\pi\lambda_{Dh}/\ell_h)^2}\left(\frac{W_{E_{r0}}}{W_{B0}}\right)
\end{equation}
It shows that the current is proportional to the ratio of energy densities in the electric, $W_{E_{r0}}=\epsilon_0E_{r0}^2/2$, and magnetic, $W_{B0}=B_0^2/2\mu_0$, fields. The last expression for the current strength can be used in Eq. (\ref{eq-bfield}) for the magnetic field generated inside the electron hole. It also serves the coefficient in the dipolar field outside the hole. 
An extreme upper saturation limit of the current strength is obtained when neglecting the ratio $N_h/N$ in the above calculation. 
\begin{eqnarray}
~~~~~~~~~I_{max}&\lesssim& \frac{2eE_{r0}N\ell_h^2}{\pi B_0}\left(1+\frac{\ell_h^2}{\pi\lambda_{Dh}^2}\right)^{-\frac{1}{2}}\times\cr 
&&\times\left\{I_2(x_h)+\sum\limits_{n=1}^\infty (-1)^nI_{2(n+1)}(x_h)\right\}
\end{eqnarray}
where we singled out the dominant term from the sum in the braces. This corresponds to the late stage of the hole when most of the electrons have been squeezed out of the hole along the magnetic field by the ever increasing magnetic field. 

\subsection{\small{Hole magnetic field $B_z(0,0)$}}
By setting the trapped electrons into motion, the electron hole thus acts as an amplifier of the magnetic field producing a field on the axis of the hole which is given by
\begin{equation}\label{eq-bfieldone}
B_z(0,0)=\frac{\mu_0NT_{eh}}{2B_0}\Big\{I_0\big[sI_0\left(x_h\right)\big]-I_0\left(s\right)\Big\}
\end{equation}
A more useful approximation for the axial field is obtained from the expansion of this expression as
\begin{equation}\label{eq-magf}
B_z(0,0)\approx \frac{B_0}{8\pi}\frac{[I_0(x_h)]^2-1}{1+(\pi\lambda_{Dh}/\ell_h)^2}\left(\frac{W_{E_{r0}}}{W_{B0}}\right)
\end{equation}
Here, the factor of $B_0$ on the right hand side is the magnetic field amplification factor. 

This field adds to the original field, $B_0$. Since it enters the azimuthal E$\times$B-drift of the magnetised hole electrons, it acts decreasing on the azimuthal current thus contributing to saturation of the magnetic field. Replacing $B_0$ in the denominator with $B_0+B_\mathit{sat}(0.0)$, one may boldly estimate a limit, $B_\mathit{sat}(0,0)$, of the saturated hole-generated magnetic field 
\begin{equation}
\frac{B_\mathit{sat}(0,0)}{B_0}={\frac{1}{2}}\Big\{\Big[\beta_{h0}\big\{I_0[sI_0(x_h)]-I_0(s)\big\}+1\Big]^\frac{1}{2}-1\Big\}
\end{equation}
where we have used the hole plasma beta $\beta_{h0}=2\mu_0NT_{eh}/B_0^2$ based on total plasma density, $N$, and ambient field, $B_0$. The final internal field then becomes
\begin{equation}
B_\mathit{max}(0.0)=B_0\left(1+B_\mathit{sat}/B_0\right) 
\end{equation}
which, when inserting for the ratio in the parentheses yields 
\begin{equation}
B_\mathit{max}(0,0)=\frac{B_0}{2}\Big\{\Big[\beta_{h0}\big\{I_0[sI_0(x_h)]-I_0(s)\big\}+1\Big]^\frac{1}{2}+1\Big\}
\end{equation}

A somewhat more convenient expression is obtained when using the above approximate expansion. This yields for the extreme upper limit, somewhat overestimating both effects, the amplification as the retardation, 
\begin{equation}\label{eq-bfieldfin}
B_\mathit{max}(0,0)\lesssim \left(\frac{\mu_0W_{Er0}}{4\pi}\right)^\frac{1}{2}I_0(x_h)
\end{equation}
Since both effects counteract each other, their actions will partially cancel such that the last expression may be taken as an estimate of the final saturation magnetic field strength in the centre of the hole.

\section{Consequences}
The generation of quasi-stationary magnetic fields inside electron holes has a profound effect on the magnetic structure of the collisionless current carrying plasma once electron holes are produced. The extension of such holes along the magnetic field is of the order of several tens of Debye lengths, and holes are generated in large quantities, each of the contributing to the amplification of the magnetic field inside and the weakening of the magnetic field outside the hole by superimposing the dipolar hole magnetic field. 
\begin{figure}[t!]
\centerline{{\includegraphics[width=0.5\textwidth,clip=]{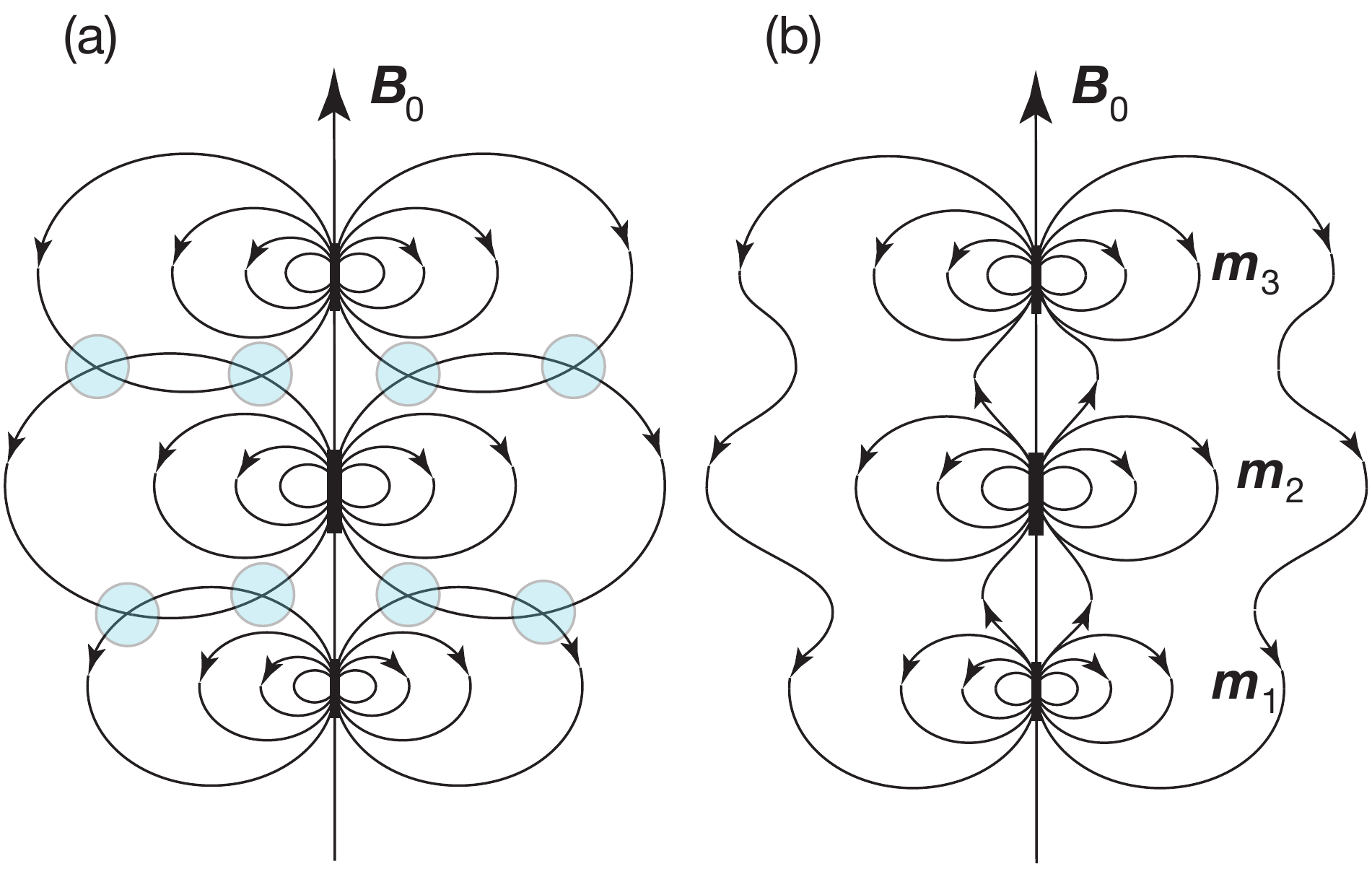}
}}\vspace{-3mm}
\caption[ ]
{\footnotesize {The external dipolar magnetic fields of three electron holes aligned on an ambient magnetic field line. The  field sources are the proper magnetic moments $\mathbf{m}_i$ generated inside the holes $(i=1,2,3)$. Along the magnetic field, the holes attract each other. \emph{(a)} The original field configuration. The blue circles indicate the sites of hole-magnetic field merging (reconnection). \emph{(b)} Magnetic field configuration caused after field line merging has occurred and fields have not completely relaxed yet.}}
\label{fig-bun-2}
\end{figure}
\subsection{\small{Magnetic structuring}}
Holes form chains of successive holes along the direction of the currents looking in simulations like collections of bracelets, each of the hole amplifying the magnetic field inside the hole and weakening the magnetic field outside the hole. Moreover, the quasi-dipolar fields of the holes in the chains contact each other. Being of antiparallel direction in their region of contact, they will readily undergo magnetic  reconnection, which is a very weak effect which probably does not contribute strongly to jetting of plasma. However, by reconnecting, the quasi-dipolar magnetic field lines of the various holes in a chain of hole will merge to become elongated flux tubes of flux that is directed oppositely to the external ambient field. This will cause a profound effect on the magnetic field. While the hole chain amplifies the field along the common current flux tube, it annihilates the ambient field outside the current carrying flux tube of hole formation. Thus the magnetic field becomes striated with strong, amplified magnetic flux tubes populated by holes and adjacent regions of weakened magnetic field. The amplification of the field takes place on a scale of $\langle x_\perp\rangle\approx 2 \langle r_h\rangle$, i.e. on the scale of the average perpendicular width of the hole chain which is caused by the Buneman two-stream wave train. 

One may expect this effect of current striation to realise in various places where sufficiently strong magnetic-field aligned currents flow. This is the case in the aurora, in particular in the downward current region where the formation of a multitude of electron holes has been well documented \citep{carlson1998,ergun1998a,ergun1998b}. Other places are the magnetospheres of the larger planets. 

Particularly interesting places are, however, found in astrophysics in relativistic shocks \citep{bykov2011} and in the relativistic foreshock regions both in the absence and in the presence of cosmic rays. In the latter case the Buneman two-stream instability will evolve in the background return current long before the famous non-resonant Bell instability sets on. The latter is made responsible for magnetic field generation and further acceleration of  cosmic rays as a macroscopic instability producing magnetic structure. Buneman electron holes in the return current flow will, on the other hand, channel the current into many filaments carrying already amplified magnetic fields on scales of the background electron gyroradii. The Bell instability thus already finds some pre-formed magnetic structuring and amplified fields which might help it growing. 
\begin{figure}[t!]
\centerline{{\includegraphics[width=0.5\textwidth,clip=]{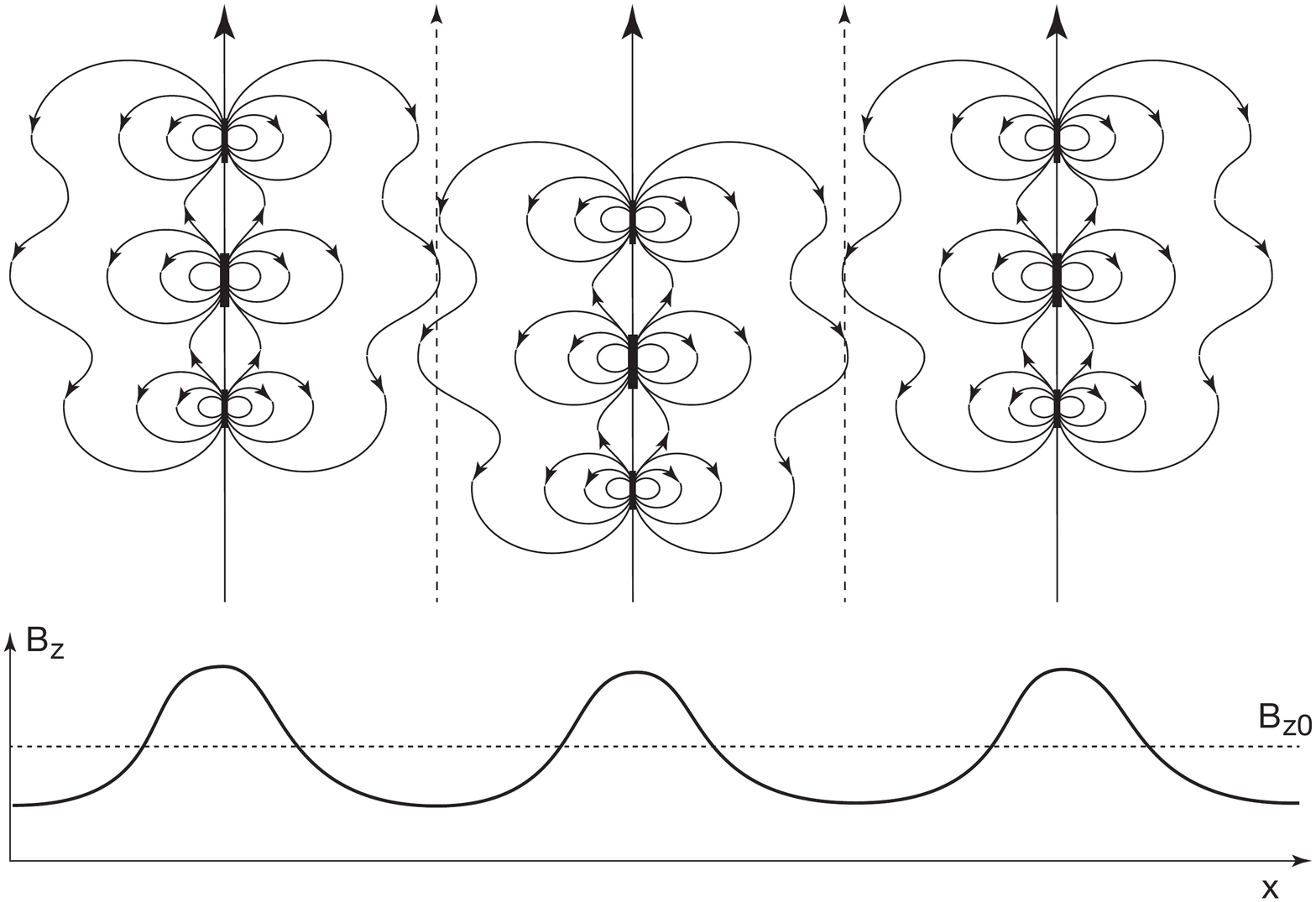}
}}
\caption[ ]
{\footnotesize {A possible spatial arrangement of holes in adjacent magnetic flux tubes in a region of laterally extended field aligned currents. The magnetic moments of nearby flux tubes tend to arrange in polar sequences ...N-S-N-S... which is partially achieved by the weakening (dotted field lines) of the external field outside the hole flux tubes caused by superposition of the antiparallel external dipolar hole field components and amplification of the field inside the hole flux tube. This alternating weakening and amplification is schematically illustrated in the lower part of the figure along a section in the direction $x$ perpendicular to the ambient field $\mathbf{B}$. The magnetic field value is oscillating in an irregular but sinusoidal way around the original value $B_{z0}$ of the ambient field. This hole arrangement causes a particular fabric in the magnetic field and the pattern of the spatial hole distribution. }}
\label{fig-bun-3}
\end{figure}
\subsection{\small{Holes as quasi-particles}}
Electron holes have been identified as being positive charges on the plasma background. With the help of the calculations of the previous sections we can estimate the value of the charges carried by a typical cylindrical hole from
Eq. (\ref{eq-holecharge}), 
where $\Phi_h(r,z)$ has been given in Eq. (\ref{eq-phihole}). Integration with respect to $z$ produces a factor $\ell_h$. Similarly the radial integral yields a factor $r_h^2I_1(x_h)/x_h$, and the hole charge becomes
\begin{equation}
q_h=2\pi\epsilon_0U\ell_h\left(\frac{r_h}{\lambda_{Dh}}\right)\left[1+\left(\frac{\pi\lambda_{Dh}}{\ell_h}\right)^2\right]^{-\frac{1}{2}}I_1(x_h)
\end{equation}
In addition to this positive charge, and electron hole carries a magnetic moment which is created by the azimuthal current $J_\phi$ flowing in the hole. This magnetic moment is responsible for the external dipolar field of the hole. It is directed along the $z$-axis of the hole, being parallel to the ambient field. Its value is
\begin{equation}
m_h=\frac{\mu_0}{4}I_\phi r_h^2
\end{equation}
It is not a genuinely conserved quantity as in the case of real particles. Electron holes thus appear as quasi-particles which carry a charge $q_h$ and a magnetic moment $m_h$.  In addition, because when moving they transport the trapped particle component along with them, they posses a mass , $M_h$, which is given by the integral over the trapped particle component in the hole
\begin{equation}
M_h=2\pi m_eN\int\limits_0^{r_h} r dr\int\limits_{-\ell_h/2}^{\ell_h/2}dz\ \exp\left[\frac{e\Phi_h}{T_{eh}}\right] 
\end{equation}
which can be brought into a more compact form by introducing the hole volume, ${\cal V}_h= 2\pi r_h^2\ell_h$, and the total number of electrons, ${\cal N}_h=N{\cal V}_h$, which would fill the volume of the hole. In this expression, the hole potential is given in Eq. (\ref{eq-phihole}). Performing the integration with respect to $z$ yields
\begin{equation}
M_h=m_e{\cal N}_h\int\limits_0^1\xi d\xi\ I_0\left\{\frac{eU}{T_{eh}}I_0\left[\xi\frac{r_h}{\lambda_{Dh}}\left(1+\frac{\pi^2\lambda_{Dh}^2}{\ell_h^2}\right)^\frac{1}{2}\right]\right\}
\end{equation}
The integral factor in this expression is smaller than one because it reduces the total number of electrons to the fraction of electrons trapped by the hole.

The three numbers, $q_h, m_h, M_h$ form a triplet which characterises the macroscopic properties of an electron hole as a heavy charged  magnetic quasi-particle. This allows to develop a kinetic theory of the interaction of electron holes in a plasma. As for an example we may consider the gyration of a hole with sim perpendicular velocity $v_{h\perp}$ in a magnetic field. Being a positive charge, the hole gyrates in the same direction as an ion, i.e. it performs a left-handed circular motion in a sufficiently strong magnetic field with cyclotron frequency and gyroradius 
\begin{equation}
\omega_{ch}=q_hB_0/M_h,\qquad r_{ch}=v_{h\perp}/\omega_{ch}
\end{equation}
In a cross-magnetic electric field $\mathbf{E}_\perp$, this charging property of the hole gives rise to an E$\times$B-drift of the hole and, in the case of many holes, of the entire hole population, causing a displacement of the hole structure across the electric and magnetic fields at the mass-independent convection velocity $V_E=E_\perp/B_0$. Holes are thus participating in this convective drift motion by being charged.

Moreover, carrying a magnetic moment, the hole will be subject to the mirror force $-\nabla_\| (m_hB_0)$ in a converging magnetic field, which will reflect or accelerate it along the magnetic field. In magnetic wells electron holes can thus become trapped around the location of minimum magnetic field strength unless pulled out by a strong external field aligned electric field $E_\|$. The parallel equation of motion of the hole in this case is
\begin{equation}
\frac{d}{dt}(M_hV_{h\|})=q_hE_\|-\nabla_\|(m_hB)
\end{equation}
where $V_{h\|}$ is the parallel hole velocity. This equation takes into account that the mass, charge and magnetic moment of the hole may change with time and location along the magnetic field. In addition, if the magnetic field changes direction during the motion of the hole along the field, then the hole magnetic moment will experience a torque
\begin{equation}
\mathbf{N}_h\approx \mathbf{m}_h\times\mathbf{B}(0)
\end{equation}
which forces the magnetic moment and hole to spin around the magnetic field. Such an effect will excite low frequency magnetic oscillations which are caused by the spinning motion of the hole magnetic moment.

\subsection{\small{Interaction of holes}}
Electron holes are generated in large numbers by the Buneman instability. One may therefore expect that they interact via the electric force acting on one electron in the external field of the other electron hole on the plasma background. If one assumes that the plasma readily neutralises the external electric field of an electron hole on the scale of the external Debye length, $\lambda_D$ (or the skin depth $\lambda_e$), then the electrostatic interaction of holes is restricted to distances $<(\lambda_D,\lambda_e)$, i.e. if the holes come into close contact. However, the holes will nevertheless interact via their magnetic moments as these cannot be compensated. In this case each hole represents a magnetic dipole of moment $\mathbf{m}_h$ directed along the external magnetic field. 

If considering only binary interactions between two nearest neighbour dipoles, $\mathbf{m}_{hj}, \mathbf{m}_{hk}$, then the potential energy of the dipolar coupling is given by \citep{jackson1975}
\begin{equation}\label{eq-dipint}
U_{jk}(r_{jk})=-\frac{\mu_0}{4\pi r_{jk}^3}\left[3(\mathbf{m}_{hj}\cdot \hat e_{jk})(\mathbf{m}_{hk}\cdot \hat e_{jk})-\mathbf{m}_{hj}\cdot\mathbf{m}_{hk}\right]
\end{equation}
where $r_{jk}$ is the distance between the centres of the two dipoles. Considering a linear chain of holes along the magnetic field, this expression reduces to
\begin{equation}
U_{jk}(r_{jk})=-\frac{\mu_0}{2\pi z_{jk}^3}m_{hj}m_{hk}
\end{equation}
where now $z_{kj}$ is the distance between two holes along $\hat z$. The magnetic force between the dipoles
\begin{equation}\label{eq-binforce}
\mathcal{F}_{jk}=-\nabla_z U_{jk} =-\frac{3\mu_0}{2\pi z^4_{jk}}\frac{\partial z_{jk}}{\partial z}m_{hj}m_{hk}
\end{equation}
as seen from an external point $z$ along the magnetic field, is attractive. The holes will attract and approach each other and will form chains of holes. When approaching closer than a distance of the screening length of the electric hole field they start feeling the repulsive electric force of the holes which on this scale competes with the attractive dipole coupling and prevents merging of the holes. Therefore, electron holes will form chains, consisting of closely spaced holes along the magnetic field. Such one-dimensional chains are similar to polymers with the difference that the magnetic field aligns them and does not permit for any folding of the chain.  

\subsection{\small{A hole-kinetic equation: Outline of theory}}
Electron holes, when being generated in large numbers, represent a gas of positively charged quasi-particles. This does, in principle, allow for a statistical dynamical description of the interaction of a gas of electron holes. In the absence of an external magnetic field, the holes would interact via the electric field of their positive charges. This case is, however, academic for two reasons. Firstly, excitation of holes via the Buneman instability requires the presence of a sufficiently strong current which necessarily has an own magnetic field. Secondly, holes are embedded in the external plasma which partially screens their proper electric field. Hence, the most probable case of hole production is along the external magnetic field in a plasma with electrostatic interactions allowed mainly along the magnetic field. 

In the presence of an external magnetic field, as our theory proves, the holes are necessarily magnetised. The single hole magnetic moments tend to align with the external magnetic field, independent on its strength. Interaction is then both electrostatic, via the screened field of the charge, and magnetic, via the magnetic moments of the holes. These interactions are binary and are thus described as collisions between the quasi-particles. 

Assuming that the quasi-particles obey a one-dimensional momentum distribution, $F_h(P_{h\|}, z,t)$ with respect to their proper magnetic field-parallel momenta, $P_{h\|}=M_hV_{h\|}$, an approximate kinetic equation describing the evolution of the hole distribution is
\begin{equation}
\frac{\partial F_h}{\partial t} +\nabla_z\left(\frac{P_{h\|}}{M_h}F_h\right) +\left[q_h{\tilde E}_\|-\nabla_z(m_hB)\right]\frac{\partial F_h}{\partial P_{h\|}}=\frac{\delta F_h}{\delta t}
\end{equation}
Here ${\tilde E}_\|=E_\|+E_{z}^\mathit{ex}$ is the sum of an external field, $E_\|$, and of the external electric field, $E_{z}^\mathit{ex}$, caused by the hole. The latter penetrates only a screening length into the plasma. This one-dimensional equation does not yet include the torque. It, however, takes into account the magnetic effect of the hole in the value of the magnetic field as given in Eqs. (\ref{eq-bfieldone}--\ref{eq-bfieldfin}). Hole plasmas are collisionless only in view of hole generation. What concerns the hole dynamics, they behave collisionally.

The interaction term on the right-hand side includes the binary magnetic dipole coupling via the force Eq. (\ref{eq-binforce}) between two nearest neighbour electron holes. This coupling is responsible for the inferred alignment of electron holes in chains and for the expected structure in the magnetic field which is caused by the large number of electron holes generated by the Buneman instability in a given field-aligned current-carrying  plasma volume. 

In the above version of the dipole coupling only linear chains of dipoles were taken into account. In a plasma volume filled with holes the structures caused will also depend on the transverse distances of the holes and of the transverse interaction between the magnetic dipoles. In this case the full form Eq. (\ref{eq-dipint}) of the binary dipole interaction potential must be used. 

The tendency of dipoles to align with their northern (N) and southern poles (S) in chains ...N-S-N-S... then causes a particular fabric of holes (see Fig. \ref{fig-bun-3}) in the Buneman unstable plasma volume which, however, will not be regular because holes have different sizes and thus possess magnetic moments of different strengths. In addition, the time variability of the holes implies a time variability of magnetic moments, which causes a time dependent magnetic fabric in the spatial region of hole generation. Such a structure highly resembles a turbulent medium. In this sense, electron holes become responsible for the generation of turbulence on the scale of the parallel and perpendicular hole dimensions. 

We should note, that similar effects are expected to arise also from ion holes where the holes represent negative charges with radially inward electric fields. The corresponding magnetic moments will be aligned anti-parallel to the magnetic field in this case, in contrast to electron holes causing weakening of the magnetic field in the interior of the holes and amplification outside. Such effects are expected to occur under weaker current conditions, when the Buneman instability is not excited and ion holes evolve from ion acoustic waves or from the modified two-stream instability. The expected magnetic field effect will then be weaker than that caused in electron holes.

\conclusions
Electron phase-space holes are three-dimensional entities. As such they, under very weak conditions on the trapped electron component, become positive charges on the electrically quasi-neutral plasma background. 

This simple fact implies that they possess a radial electric field component which, together with the ambient magnetic field, forces the trapped electron component into azimuthal motion. By virtue of the electron current  a magnetic moment is attached to the electron phase space hole.  This process has been demonstrated analytically by solving the exact boundary value problem for an cylindrical electron hole. We have determined the saturation amplitude of the secondary magnetic field generated by the hole.  

This process is of interest, less for the mere effect of electron holes being magnetised but for the resulting effect that the presence of many holes in a field-aligned current flux tube (for instance in the auroral source region or at the solar flare site) amplifies the magnetic field in the flux tube and weakens the magnetic field in the adjacent hole environment. The resulting magnetic filamentation might be of substantial interest in the dynamics of other effects like the generation of radiation, turbulence, and particle acceleration, in particular the acceleration of electrons. Since it is expected that collisionless  reconnection regions are strong sources of field aligned currents, the resulting filamentary structure may as well have consequences for the secondary effects of collisionless reconnection, structuring of fields, acceleration of particles, and turbulence.

Another important application would be the foreshock region of relativistic shocks \citep{bykov2011}. Since the work of \citet{bell2004} it is believed that relativistic parallel or quasi-parallel shocks, being surrounded by a cloud of medium energy cosmic rays, possess an extended foreshock which carries a return current flowing in the non-cosmic ray background plasma along the magnetic field and connecting the shock current with the cosmic ray cloud. These currents have been assumed to drive the non-resonant Bell instability which is believed \citep{bykov2011a} to be the source of low frequency magnetic fields which grow on the expense of the return current and cosmic ray energy and are required for making the relativistic shock an efficient accelerator for cosmic rays. In order to do this, the currents must be strong, in which case the Buneman instability will necessarily be excited long before the Bell instability starts growing. This will cause a large number of electron holes to evolve in the relativistic foreshock, causing generation of small scale magnetic fields which amplify the foreshock magnetic field and structure it into filaments. When this happens, the relativistic foreshock makes the transition to turbulence on the scales of the electron phase-space hole gas. The Bell instability then starts at much later times, when hole formation has long saturated, from this turbulent state and magnetic filaments in the foreshock, which might help it growing to larger than so far estimated magnetic field amplitudes.     

We have not yet exploited the effects of the torque which will necessarily be generated when the hole passes along a magnetic field changing direction. This is the case, for instance, in the auroral region over scales of a fraction of an Earth radius, the order of between 100 km and 1000 km. Holes are believed to survive over such scales. The torque exerted on the hole magnetic moment by the spatial change of the magnetic field direction causes nutation of the hole around the magnetic field, thus giving rise to a magnetic oscillation with frequency of the hole nutation. Oscillations of this kind, if produced in regions of large hole density and comparable magnetic moments, will necessarily give rise to irregular low-amplitude low-frequency magnetic field variations of life time of the average life time of electron holes, possibly detectable in the Pi range. Similar pulsation can be expected, for instance, in solar flare regions where they might be responsible for pulsations in the observed radio emission.

\begin{acknowledgements}
This research was part of an occasional Visiting Scientist Programme in 2006/2007 at ISSI, Bern. 
RT  thankfully recognises the assistance of the ISSI librarians, Andrea Fischer and Irmela Schweizer. He appreciates the encouragement of Andr\'e Balogh, Director at ISSI. 
\end{acknowledgements}


\begin{thebibliography}{ }


\bibitem[Andersson et al.(2009)]{andersson2009} Andersson, L., Ergun, R. E., Tao, J., Roux, A., LeContel, O., Angelopoulos, V., Bonnell, J., McFadden, J. P., Larson, D. E., Eriksson, S., Johansson, T., Cully, C. M., Newman, D. L., Goldman, M. V., Glassmeier, K.-H., \& Baumjohann, W.: New features of electron phase space holes observed by the THEMIS mission, Phys. Rev. Lett. 102, 225004, doi: 10.1103/PhysRevLett.102.225004, 2009. 

\bibitem[Bale et al.(1998a)]{bale1998a} Bale, S. D., Kellogg, P. J., Larson, D. E., Lin, R. P., Goetz, K. \& Lepping, R. P.: Bipolar electrostatic structures in the shock transition region: Evidence of electron phase space holes, Geophys. Res. Lett. 25, 2929, doi: 10.1029/98GL02111, 1998a. 

\bibitem[Bale et al.(1998b)]{bale1998b} Bale, S. D., Kellogg, P. J., Goetz, K. \& Monson, S. J.: Transverse z-mode waves in the terrestrial electron foreshock, Geophys. Res. Lett. 25, 9-12, doi: 10.1029/98GL03493, 1998b. 

\bibitem[Bale et al.(2002)]{bale2002} Bale, S. D., Hull, A., Larson, D. E., Lin, R. P., Muschietti, L., Kellogg, P. J., Goetz, K. \& Monson, S. J.: Electrostatic turbulence and Debye-scale structures associated with electron thermalization at collisionless shocks, Astrophys. J. Lett. 575, 9L25-L28, doi: 10.1086/342609, 2002. 


\bibitem[Baumjohann and Treumann(2012)]{baumjohann2012} Baumjohann, W. \& Treumann, R. A.: Basic Space Plasma Physics (Revised Edition), Imperial College Press, London 2012, chpt. 11. 


\bibitem[Bell(2004)]{bell2004} Bell, A. R.: Turbulent amplification of magnetic field and diffusive shock acceleration of cosmic rays, Monthly Notic. Royal Astron. Soc. 353, 550, doi: 10.1111/j.1365-2966.2005.08774.x, 2004.

\bibitem[Bell(2005)]{bell2005} Bell, A. R.: The interaction of cosmic rays and magnetized plasma, Monthly Notic. Royal Astron. Soc. 358, 181, doi: 10.1111/j.1365-2966.2004.08097.x, 2005.

\bibitem[Bernstein et al.(1957)]{bernstein1957} Bernstein, I. B., Greene, J. M.  \& Kruskal, M. D.: Exact nonlinear plasma oscillations, Phys. Rev. 108, 546-550, doi: 10.1103/PhysRev.108.546, 1957.

\bibitem[Bret(2009)]{bret2009} Bret, A.: Weibel, two-stream, filamentation, oblique, Bell, Buneman...Which one grows faster?, Astrophys. J. 699, 990-1003, 2009. 

\bibitem[Bret(2011)]{bret2011} Bret, A.: Rigorous merging of two-stream and Buneman instabilities, Phys. Scripta 84, ID 065507, 2011. 

\bibitem[Buneman(1958)]{buneman1958} {Buneman, O.}: Instability, turbulence, and conductivity in current-carrying plasma, Phys. Rev. Lett. 1,  8-9, 1958, doi: 10.1103/PhysRevLett.1.8.

\bibitem[Buneman(1959)]{buneman1959} {Buneman, O.}: Dissipation of currents in ionized media, Phys. Rev. 115, 503-517, 1959, doi: 10.1103/PhysRev.115.503.

\bibitem[Bykov et al.(2011)]{bykov2011a} Bykov, A. M., Osipov, S. M. \& Ellison, D. C.: Cosmic ray current driven turbulence in shocks with efficient particle acceleration: the oblique, long-wavelength mode instability, Monthly Notic. Royal Astron. Soc. 410, 39, doi: 10.1111/j.1365-2966.2010.17421.x, 2011.


\bibitem[Bykov and Treumann(2011)]{bykov2011} Bykov, A. M. \& Treumann, R. A.: Fundamentals of collisionless shocks for astrophysical application: Relativistic shocks, Astroph. Astron. Rev. 19, ID 42, doi: 10.1007/s00159-011-0042-8, 2011. 

\bibitem[Carlson et al.(1998)]{carlson1998} Carlson, C.~W.,  McFadden, J. P., Ergun, R. E., Temerin, M., Peria, W., Mozer, F. S., Klumpar, D. M., Shelley, E. G., Peterson, W. K., Moebius, E., Elphic, R., Strangeway, R., Cattell, C. \& Pfaff, R.: FAST observations in the downward auroral current region: Energetic upgoing electron beams, parallel potential drops, and ion heating, Geophys. Res. Lett. 25, 2017-2020, doi:  	10.1029/98GL00851, 1998.



\bibitem[Chen and Parks(2002)]{chen2002} Chen, L.  \& Parks, G. K.: BGK electron solitary waves in 3D magnetized plasma, Geophys. Res. Lett. 29, 1331, doi: 10.1029/2001GL013385, 2002.

\bibitem[Chen et al.(2004)]{chen2004} Chen, L., Thouless, D. J.  \& Tang, J.: Bernstein Greene Kruskal solitary waves on three-dimensional magnetized plasma, Phys. Rev. E 69, 055401, doi: 10.1103/PhysRevE.69.055401, 2004.









\bibitem[Drake et al.(2003)]{drake2004} Drake, J., Swisdak, M., Cattell, C., Shay, M. A., Rogers, B. N. \& Zeiler, A.: Formation of electron holes and particle energization during mgnetic reconnection, Science 299, 837-877, doi: 10.1126/science.1080333, 2003.

\bibitem[Du et al.(2011)]{du2011} Du, A., Wu, M., Lu, Q., Huang, C. \& Wang, S.: Transverse instability and magnetic structures associated with electron phase space holes, Phys. Plasmas 19, 159-162, ID 032104, doi: 10.10263/1.3561796, 2011.

\bibitem[Dum(1989)]{dum1989} Dum, C.~T.: Transition in the dispersive properties of beam-plasma and two-stream instabilities, J. Geophys. Res. 94, 2429-2442, doi: 10.1029/JA094iA03p02429, 1989. 

\bibitem[Dum and Nishikawa(1994)]{dum1994} Dum, C.~T. \& Nishikawa, K.-I.: Two-dimensional simulation studies of the electron beam-plasma instability, Phys. Plasmas 1, 1821-1826, doi: 10.1063/1.870636, 1994. 


\bibitem[Dupree(1972)]{dupree1972} Dupree, T.~H.: Theory of phase space density granulation in plasma, Phys. Fluids 15, 334-344, doi: 10.1063/1.16939111972. 


\bibitem[Dupree(1982)]{dupree1982} Dupree, T.~H.: Theory of phase-space density holes,  Phys. 
Fluids 25, 277-289, doi: 10.1063/1.863734, 1982. 

\bibitem[Dupree(1983)]{dupree1983} Dupree, T.~H.: Growth of phase-space density holes, Phys. 
Fluids 26, 2460-2481, doi: 10.1063/1.864430, 1983. 



\bibitem[Ergun et al.(1998a)]{ergun1998a}
Ergun, R.~E., {Carlson, C. W., McFadden, J. P., Mozer, F. S., Delory, G. T., Peria, W., Chaston, C. C., Temerin, M., Roth, I., Muschietti, L., Elphic, R., Strangeway, R., Pfaff, R., Cattell, C. A., Klumpar, D., Shelley, E.,  Peterson, W.,  Moebius, E., \& Kistler, L.}: FAST satellite observations of large-amplitude solitary
structures, Geophys. Res. Lett., 25, 2041--2044, {doi:10.1029/98GL00636}, 1998a.

\bibitem[Ergun et al.(1998b)]{ergun1998b}
Ergun, R.~E., {Carlson, C. W., McFadden, J. P., Mozer, F. S., Muschietti, L., Roth, I. \& Strangeway, R. J.}: Debye-scale plasma structures associated with
magnetic-field-aligned electric fields, Phys. Rev. Lett., 81, 826--829, {doi:10.1103/PhysRevLett.81.826}, 1998b.

\bibitem[Ergun et al.(1998c)]{ergun1998c} 
Ergun, R.~E., {Carlson, C. W., McFadden, J. P., Mozer, F. S., Delory, G. T., Peria, W., Chaston, C. C., Temerin, M., Elphic, R., Strangeway, R., Pfaff, R., Cattell, C. A., Klumpar, D., Shelley, E.,  Peterson, W.,  Moebius, E. \& Kistler, L.}: FAST satellite observations of electric field structures in the auroral zone, Geophys. Res. Lett. 25, 2025-1028, {doi: 10.1029/98GL00635}, 1998c.

\bibitem[Ergun et al.(2002a)]{ergun2002a}
{Ergun, R.~E., Andersson, L., Main, D., Su, Y.-J., Carlson, C. W., McFadden, J. P. \& Mozer, F. S.: Parallel electric fields in the upward current region of
the aurora: Indirect and direct observations, Phys. Plasmas, 9, 3685--3694, {doi:10.1063/1.1499120}, 2002a.}

\bibitem[Ergun et al.(2002b)]{ergun2002b}
Ergun, R.~E., {Andersson, L., Main, D., Su, Y.-J., Newman, D. L., Goldman, M. V., Carlson, C. W., McFadden, J. P. \& Mozer, F. S.}: Parallel electric fields in the upward current region of
the aurora: Numerical solutions, Phys. Plasmas, 9, 3695--3704, {doi:10.1063/1.1499121}, 2002b.



\bibitem[{{Franz et al.}(2000)}]{franz2000}{Franz} J. R., Kintner, P. M., Seyler, C. E., Pickett, J. S. \& Scudder, J. D.: {On the perpendicular scale of electron phase-space holes}, Geophys. Res. Lett. 27, 169-1672, doi: {10.1029/1999GL010733}, 2000.




\bibitem[Jackson(1975)]{jackson1975} Jackson, J. D.: Classical Electrodynamics (Second Edition), John Wiley \& Sons, Inc., New York, 1975, chp. 5.



\bibitem[Jovanovic and Horton(1993)]{jovanovic1993} Jovanovic, D.  \& Horton, W.: Quasi-three-simensional electron holes in magnetized plasmas, Phys. Fluids B5, 433-439, doi: 10.1063/1.860528, 1993.


















\bibitem[Matsukiyo et al.(2004)]{matsukiyo2004} Matsukiyo, S., Treumann, R.~A. \& Scholer, M.: Coherent waveforms in the auroral upward current region, J. Geophys. Res. 109, ID A06212, doi: 10.1029/2004JA010477, 2004. 

\bibitem[Mozer et al.(2002)]{mozer2002} Mozer, F. S., Bale, S. D. \& Phan, T. D.: Evidence of diffusion regions at a subsolar magnetopause crossing, Phys. Rev. Lett. 89, ID 015002, doi: 10.1103/PhysRevLett.89.015002, 2002. 

\bibitem[Mozer et al.(2008)]{mozer2008} Mozer, F. S., Angelopoulos, V., Bonnell, J., Glassmeier, K.-H. \& McFadden, J. P.: Evidence of diffusion regions at a subsolar magnetopause crossingTHEMIS observations of modified Hall fields in an asymmetric magnetic field reconnection, Geophys. Res. Lett. 35, ID L17S04, doi: 10.1029/2007GL033033, 2008. 



\bibitem[Mozer and Pritchett(2010)]{mozer2010} Mozer, F. S. \& Pritchett, P. L.: Spatial, temporal, and amplitude characteristics of parallel electric fields associated with subsolar magnetic field reconnection, J. Geophys. Res. 115, ID A04220, doi: 10.1029/2009JA014718, 2010. 


\bibitem[Mozer and Pritchett(2011)]{mozer2011} Mozer, F. S. \& Pritchett, P. L.: Electron physics of asymmetric magnetic field reconnection, Space Sci. Rev. 158, 119-143, doi: 10.1007/s11214-010-9681-8, 2011. 


\bibitem[Muschietti et al.(1999a)]{muschietti1999a} Muschietti, L., Roth, I., Ergun, R.~E. \& Carlson, C.~W.: Analysis and simulation of BGK electron holes, Nonlin. Process. Geophys. 6, 211-219, 1999a. 

\bibitem[Muschietti et al.(1999b)]{muschietti1999b} Muschietti, L., Ergun, R.~E., Roth, I. \& Carlson, C.~W.: Phase-space electron holes along magnetic field lines, Geophys. Res. Lett.  26, 1093-1096, doi: 10.1029/1999GL900207, 1999b.

\bibitem[Newman et al.(2001)]{newman2001} Newman, D.~L., Goldman, M.~V., Ergun, R.~E. \& Mangeney, A; Formation of Double Layers and Electron Holes in a Current-Driven Space Plasma, Phys. Rev. Lett. 87, 255001, doi: 10.1103/PhysRevLett.87.255001, 2001. 

\bibitem[Newman et al.(2002)]{newman2002} Newman, D.~L., Goldman, M.~V. \& Ergun, R.~E.: Evidence for correlated double layers, bipolar structures, and very-low-frequency saucer generation in the auroral ionosphere, Phys. Plasmas 9, 2337-2343, doi: 10.1063/1.1455004, 2002.

\bibitem[Oppenheim et al.(2001)]{oppenheim2001} {Oppenheim, M. M., Vetoulis, G.,  Newman, D.~L. \&  Goldman, M.~V.: Evolution of electron phase-space holes in 3D, Geophys. Res. Lett. 28, 1891--1894, doi: 10.1029/2000GL012383, 2001.}

\bibitem[Pickett et al.(2004)]{pickett2005} Pickett, J., Chen, L., 
Kahler, S., Santolik, O., Gurnett, D., Tsurutani, B. \& Balogh, A.: 2004, Isolated electrostatic structures observed throughout the Cluster orbit: Relationship top magnetic field strength, Ann. Geophys. 22, 2515-2523, doi: 10.5194/angeo-22-2515-2004, 2004. 

\bibitem[Pottelette and Treumann(2005)]{pottelette2005} Pottelette, R. \& Treumann, R. A.: Electron holes in the auroral upward current region,  Geophys.  Res. Lett. 1432, L12104, doi: 10.1029/2005GL022547, 2005.

\bibitem[Schamel(1975)]{schamel1975} Schamel, H.: Analytic BGK modes and their modulational instability, J. Plasma Phys. 13, 139-145, doi: 10.1017/S0022377800025927, 1975.

\bibitem[Schamel(1979)]{schamel1979} Schamel, H.: 1979, Theory of electron holes, Phys. Scripta 
20, 336-342, doi: 10.1088/0031-8949/20/3-4/006, 1979. 

\bibitem[Schamel and Bujarbarua(1983)]{schamel1983} Schamel, H., and Bujarbarua, S.:  Analytical double layers, Phys. Fluids 26, 190-193, doi: 10.1063/1.864006, 1983. 


\bibitem[Schamel(1986)]{schamel1986} Schamel, H.: Electron holes, ion holes and double layers: Electrostatic phase space structures in theory and experiment, Phys. Reports 140, 161-191, doi: 10.1016/0370-1573(86)90043-8, 1986.












\bibitem[Tao et al.(2011)]{tao2011} Tao, J. B., Ergun, R. E., Andersson, L., Bonnell, J. W., Roux, A., LeContel, O., Angelopoulos, V., McFadden, J. P., Larson, D. E., Cully, C. M., Auster, H.-U., Glassmeier, K.-H., Baumjohann, W., Newman, D. L. \& Goldman, M. V.: A model of electromagnetic electron phase-space holes and its application, J. Geophys. Res. 116, A11213, doi: 10.1029/2010JA016054, 2011.

\bibitem[Terry et al.(1990)]{terry1990} Terry, P. W., Diamond, P. H.  \& Hahm, T. S.: The structure and dynamics of electrostatic and magnetostatic drift holes, Phys. Fluids B2, 2048-2063, doi: 10.1063/1.859426, 1990.


\bibitem[Treumann(2006)]{treumann2006} Treumann, R. A.: The electron-cyclotron maser for astrophysical application, Astron. Astrophys. Rev. 13, 229-315, doi: 10.1007/s00159-006-0001-y, 2006.

\bibitem[Treumann(2009)]{treumann2009} Treumann, R. A.: Fundamentals of collisionless shocks for astrophysical application, 1. Non-relativistic shocks, Astron. Astrophys. Rev. 17, 409-535, doi: 10.1007/s00159-009-0024-2, 2009.

\bibitem[Treumann et al.(2012)]{treumann2012} Treumann, R.~A., 
Baumjohann, W. \& Pottelette, R.: Electron-cyclotron maser radiation from electron holes: Downward current region, Ann. Geophys. 30, 119-130, doi:  10.5194/angeo-30-119-2012.



\bibitem[Turikov(1984)]{turikov1984} Turikov, V. A.: Electron phase space holes as localized BGK solutions, Phys. Scripta 30, 73, doi: 10.1088/0031-8949/30/1/015, 1984.




\bibitem[Umeda et al.(2004)]{umeda2004} Umeda, T., Omura, Y. \& Matsumoto, H.: Two-dimensional particle simulations of electromagnetic field signature associated with electrostatic solitary waves, J. Geophys. Res.  109, A02207, doi: 10.1029/2003JA010000, 2004.



\bibitem[Umeda et al.(2006)]{umeda2006} Umeda, T., Omura, Y, Miyake, T. Matsumoto, H.  \& Ashour-Abdalla, M.: Nonlinear evolution of the electron two-stream instability: Two-dimensional particle simulations, J. Geophys. Res.  111, A10206, doi: 10.1029/2006JA011762, 2006.

\bibitem[Vaivads et al.(2004)]{vaivads2004} Vaivads, A., Khotyaintsev, Y., Andr{\'e}, M., Retin\`o, A., Buchert, S. C., Rogers, B. N., D\'ecr\'eau, P., Paschmann, G. \& Phan, T. D.:  Structure of the magnetic reconnection diffusion region from four-spacecraft observations, Phys. Rev. Lett. 93, ID 105001, doi:  10.1103/PhysRevLett.93.105001, 2004.


\bibitem[Weibel(1959)]{weibel1959} Weibel, E. S.:  Spontaneously growing transverse waves in a plasma due to an anisotropic velocity distribution, Phys. Rev. Lett. 2, 83-84, doi: 10.1103/PhysRevLett.2.83, 1959.

\bibitem[Wu et al.(2011)]{wu2011} Wu, M., Lu, Q., Du, A., Xie, J. \& Wang, S.: The evolution of the magnetic structures in electron phase-space holes: Two-dimensional particle-in-cell simulations, J. Geophys. Res. 116, A10208, doi:10.1029/2011JA016486, 2011.






\end{thebibliography}
\end{document}